\documentclass[psfig]{aa}                                                    
\input psfig.tex

\begin{document}                                                                
\def\et{et al.}                                                                 
\def\egs{erg s$^{-1}$}                                                          
\def\egsc{erg s$^{-1}$ cm$^{-2}$}                                               
\def\msu{M$_{\odot}$\ }                                                         
\def\kms{km s$^{-1}$ }                                                          
\def\kmsM{km s$^{-1}$ Mpc$^{-1}$ }                                              
                                                                                
   \title{The New Emerging Model for the Structure of Cooling Cores in
 Clusters of Galaxies}

   \titlerunning{Galaxy Cluster Cooling Cores}

   \author{H. B\"ohringer \inst{1}, K. Matsushita \inst{1},
   E. Churazov \inst{2}, Y. Ikebe, \inst{1}, Y. Chen \inst{1}} 
                                                                                
   \authorrunning{B\"ohringer et al.}
                                             
   \offprints{H. B\"ohringer \\ hxb@mpe.mpg.de}                                 
                                                                                
   \institute{$^1$ Max-Planck-Institut f\"ur Extraterrestrische Physik,         
                   D-85748 Garching, Germany\\                                  
              $^2$ Max-Planck-Institut f\"ur Astrophysik,         
                   D-85748 Garching, Germany\\
                    } 
                                                                                
   \date{Received ..... ; accepted ...}
                                                                                
   \markboth {Galaxy Cluster Cooling Cores}{}                                  
                                                                                
\abstract{                                                                
New X-ray observations with XMM-Newton show a lack of spectral evidence
for large amounts of cooling and condensing gas in the centers of 
galaxy clusters believed to harbour strong cooling flows.
The paper reexplores the cooling flow scenario in the light of the new 
observations. We explore the diagnostics of the temperature structure 
of cooling cores with XMM-spectroscopy, tests for intracluster X-ray 
absorption towards central AGN, the effect of metal abundance inhomogeneities,
and the implications of high resolution images in the centers of clusters. 
We find no evidence of intrinsic absorption in the center of the 
cooling flows of M87 and the Perseus cluster. We further
consider the effect of cluster rotation in cooling flow regions in
the frame of cosmic structure evolution models.
Also the heating of the core regions of clusters by jets from a central 
AGN is reconsidered. We find that the power of the AGN jets as
estimated by their interaction effects with the intracluster medium
in several examples is more then sufficient to heat the cooling flows
and to reduce the mass deposition rates. We explore in more detail which
requirements such a heating model has to fulfill to be consistent
with all observations, point out the way such a model could be 
constructed, and argue that such model building seems to be successful.  
In summary it is argued that most observational
evidence points towards much lower mass deposition rates 
than previously inferred in the central
region of clusters thought to contain strong cooling flows.}
      
\maketitle
                                                                                
%
                                                                                
\section{Introduction}                         

With first detailed X-ray observations using the {\sl UHURU} and
{\sl Copernicus} satellites and the first rocket borne X-ray telescopes
it was discovered that the X-ray emitting, hot gas in galaxy clusters
reaches high enough densities in the cluster centers that the cooling
time of the gas falls below the Hubble time. The consequences of these
observations have first been explored in  early papers by
Silk (1976), Fabian \& Nulsen (1977), Cowie \& Binney (1977), and Mathews
and Bregman (1978). In the absence of a suitable fine-tuned heating source,
the cooling and condensation of the gas in the central regions 
is a straight-forward consequence of the energy equation of the hot
gas. From this analysis the cooling flow scenario emerged (e.g. Fabian
et al. 1984, Fabian 1994). This scenario obtained very early strong 
support from spectroscopic observations with the {\sl Einstein} 
observatory {\sl Solid State Spectrometer} and {\sl Focal Plane
Cristal Spectrometer} instruments.
It was found that the spectra imply an emission measure distribution
of the gas as a function of temperature showing low temperature phases
in addition to the hotter temperature gas and a distribution function
in temperature consistent with steady state cooling (Canizares et al. 1979,
1982, Mushotzky \& Szymkowiak 1988). This spectroscopic diagnostics 
remained one of the strongest bases for the foundation of the cooling 
flow picture until now, where {\sl XMM-Newton} and {\sl Chandra} are 
finally providing high resolution X-ray spectra superseeding those 
obtained with the {\sl Einstein} instruments.

Detailed analysis of surface brightness profiles of X-ray images of 
clusters obtained with the {\sl Einstein} observatory and further theoretical 
work led to the detailed, self-consistent scenario of inhomogeneous, comoving
cooling flows (Nulsen 1986, Thomas, Fabian, \& Nulsen 1987). The main 
assumptions on which this model is based and some important 
implications are: (i) Each radial zone
in the cooling flow region comprises different plasma phases
covering a wide range of temperatures. The consequence
of this temperature distribution is that gas will cool to low 
temperature and condense over a wide range of radii. This is in 
contrast to a homogeneous cooling flow where the hot plasma would only 
condense and be deposited in the center, implying a very
peaked surface brightness profile which is not observed
(e.g. Fabian 1994). (ii) The
gas comprising different temperature phases features an inflow in
which all phases move with the same flow speed, forming a comoving cooling flow.
(iii) There is no energy exchange between the different phases,
between material at different radii, and no heating. This means in 
particular that heat conduction has to be suppressed to a negligible 
value (e.g. Fabian, Canizares \& B\"ohringer 1994). When all this premisses
are taken together, the inhomogeneity of the cooling flow is a direct
consequence of the observed X-ray surface brightness profile 
as observed for those clusters of galaxies which feature a low 
central cooling time (Thomas et al. 1987).

Further X-ray observations 
brought relatively little new support to this detailed 
scenario in the last about 15 years. {\sl ROSAT} observations
showed that the central regions in cooling flow clusters have
a lower temperature than the outer regions (e.g. Schwarz et al. 1992). 
Spectroscopic studies with {\sl ASCA} provided evidence that there is at
least more than one temperature phase present in cluster centers 
(e.g. Fabian et al. 1994, Ikebe et al. 1997, 1999, Makishima et al. 2001). 
But these spectroscopic data obtained with {\sl ASCA} are not sufficient 
to allow a detailed test of the cooling flow model
by the reproduction of the predicted emission measure distribution
as a function of temperature as predicted for the inhomogeneous cooling
flow model (e.g. Johnstone et al. 1992). One further interesting 
observational evidence was the spectral signature of excess absorption 
in addition to the absorption expected for the interstellar medium 
of our galaxy (e.g. White et al. 1991). This excess absorption 
originates most probably in the galaxy cluster centers. It
was taken as further support for the cooling flow model since
it provides evidence that gas has cooled from the 
cooling flow to low temperature (e.g. Fabian 1994, Allen et al. 2001).

It remained a puzzle, however, that for the regions with cooling flows
with estimated mass deposition rates up to several hundred or thousand
solar masses per year little evidence for such massive
cooling of gas could be found at other
wavelength (e.g. Fabian 1994). This is a major reason why the cooling
flow model is not accepted unisonously among astronomers.  
Evidence for warm gas (diffuse emission line systems), 
cool gas, and even star formation was
well detected at smaller levels (e.g. Heckman et al. 1989, Fabian 1994).
In a recent review of indications of star formation in cooling flows
McNamara (1997) show that the inferred star formation rates are
typically only in the range 1 to 10\% of the model predicted 
mass deposition rates and range from 1 to $< 100$ M$_{\odot}$ yr$^{-1}$
(e.g. McNamara \& O'Connell 1989, 1993, Crawford \& Fabian 1992,
Allen 1995, Cardiel et al. 1998,
Crawford et al. 1999). Very recently interesting observations of 
molecular emission lines from CO and indications of warm dust
have lead to the implications of a larger reservoir of cold gas
in cooling flows (Wilman et al. 2000, Edge 2001, Edge et al. 1999).
These observations provide a stronger support that mass deposition may
happen in the more massive cooling flows, but they do not necessarily
confirm the very high mass deposition rates inferred for these clusters
from X-ray imaging data.

Therefore there was little means from observational data to prove 
or disprove the existence of inhomogeneous cooling flows with massive mass
deposition rates and one was awaiting better opportunities 
for more detailed X-ray spectroscopic observations to further test
the cooling flow picture. The Japanese {\sl ASTRO E} would have been
a prime instrument for such studies, but very unfortunately it was
lost in the launching accident. Now the first analysis of high resolution
X-ray spectra and imaging spectroscopy obtained with {\sl XMM-Newton} 
has supplied us with the first surprises which could lead to a breakthrough in the 
understanding of this long-standing debate. The first surprise
is that the spectra show no signatures of cooler phases of the 
cooling flow gas below an intermediate temperature which constitutes a 
problem for the interpretation of the results in the conventional
cooling flow picture (e.g. Peterson et al. 2001, Tamura et al. 2001). 
Another result is that the spectroscopic data are better explained 
with local isothermality in the cooling flow region (e.g. Matsushita
et al. 2001, Molendi \& Pizzolato 2001) which is also in conflict with the
inhomogeneous cooling flow model. 

In this paper we discuss these new spectroscopic results and their
implications, review some further unsolved issues concerning cooling flows,
and point out the way to a new possible model for this phenomenon.
Section 2 is concerned with the spectroscopic evidence
of the absence of low temperature phases and the local isothermality
in cooling flow regions. We take examples for the results mainly from
the best observed case, the X-ray halo of M87 
in the X-ray emission center of the
Virgo cluster. In section 3 and 4 we discuss the evidence of internal 
absorption in cooling flows in the light of new results using the
AGN in the center of clusters as independent probes. Section 5 is devoted
to a discussion of surface brightness inhomogeneities observed in 
cooling flow regions as possible sites of mass deposition. 
In section 6 we review the problem of cooling flows in rotating clusters
with respect to the expectation values for angular momentum in clusters
in the frame of cosmological models. In section 7 we discuss the requirements
for a suitable heating model and explore if such a heating model can be
devised on the basis of the energy input from AGN in the cluster centers. 
In section 8
we address the implications of the strong observed magnetic fields
in cooling flow regions and section 9 provides the conclusions.
Since we argue for strongly reduced mass condensation rates which are
more localized in the very central regions,
we will term the cooling flow regions 
``cluster cooling cores'' in the following. 

For the physical parameters of galaxy clusters that scale with distance
we use a Hubble constant of $H_0 = 50$ km s$^{-1}$ Mpc$^{-1}$.         
                                                                                
\section{Spectroscopic Evidence for the Absence of Low Temperature
Phases and Local Isothermality in Cooling Flow Regions}                      

Recent results from XMM-Newton (Jansen et al. 2001) 
observations have provided unpreceded
detailed spectroscopic diagnostics of the central regions of clusters
providing new insights into the structure of cooling core regions.
One set of observations is obtained with the XMM Reflection Grating 
Spectrometer (RGS, den Herder et al. 2001). They show for several
cooling core regions spectral signatures of different temperature
phases ranging approximately from the hot virial temperature of
the cluster to a lower limiting temperature, $T_{low}$, 
which is still significantly above the ``drop out'' temperature 
where the gas would cease to emit significant X-ray radiation. 
That is, the clearly observable spectroscopic features of the
lower temperature gas expected for a cooling flow model are
not observed. In the case of the massive, cooling flow cluster
A1835 with a bulk temperature of about 8.3 keV no lines or features for
temperature phases below 2.7 keV are found (Peterson et al. 2001). 
Similar results have been derived for the cluster A1795 
(Tamura et al. 2001).

In the standard cooling flow model the spectral luminosity from the 
different temperature phases is fixed by the mass flow rate of the cooling
flow in such a way that the emission measure for each temperature
interval, $[T , T+dT]$, from the hottest temperatures to the 
``drop out'' temperature is proportional to the mass flow rate, $\dot M$,
and given by (e.g. Johnstone et al. 1992)

\begin{equation}
L_{\nu}(T)~ dT ~~ = ~~ {\Lambda_{\nu}(T) \over \Lambda_{cool}(T)}
~{5 k_B \over 2 \mu m_p}~ \dot M~ dT ,
\end{equation}

where $\Lambda_{\nu}(T)$ and $\Lambda_{cool}(T)$ are the emissivity
at frequency $\nu$ and the bolometric emissivity, respectively, 
and $\mu m_p$ is the mean particle
mass in the hot plasma. The total spectrum is then given by the 
integral

\begin{equation}
L_{\nu}~~  = {5 k_B \over 2 \mu m_p}~ \dot M~ 
\int_0^{T_{max}} {\Lambda_{\nu}(T) \over\Lambda_{cool}(T)} dT .
\end{equation}

Therefore, if the mass deposition rate determined from the surface
brightness distribution is known, the spectral signature of each
temperature phase of the cooling flow is fixed and the total spectrum
can be predicted. That the expected low temperature signatures
are not observed is therefore in clear conflict with the standard
cooling flow model.

These results coming from the {\sl XMM RGS} data are very well confirmed
by {\sl XMM-Newton} observations  
with the energy sensitive imaging devices, EPN
(Str\"uder et al. 2001) and EMOS (Turner et al. 2001). Even though
the spectral resolution is less for these instruments than for
the RGS they can very well be used to detect temperature sensitive
spectral features with high accuracy due to the good photon statistics.
They have the additional advantage of providing spectral
information across the entire cooling core region.    
First the spectral fitting of the X-ray 
emission in the cooling flow clusters M87, A1795, A496, and A1835
showed that single temperature models provided a better representation
of the data than cooling flow models (B\"ohringer et al. 2001,
Molendi \& Pizzolato 2001) also implying the lack of low temperature
components. In a recent, very detailed analysis of the temperature
structure of the closest and best studied cooling flow in M87,
Matsushita et al. (2001) have shown that the temperature
structure is well described locally by a single temperature
over most of the cooling core region. In particular the spectral 
features originating from gas below a temperature of about 0.8 - 1 keV,
as predicted by the standard cooling flow model through the above formula,
are not detected. The approximate local isothermality of the gas
is shown very well by the overall consistency of the fitting of 
complete spectra as well as by the 
study of special features as e.g. the spectral feature of the blend 
of Fe L-shell lines and the ratio of the hydrogen like to helium like 
emission lines in Si and S. An exception is the central region 
($r \le 1$ arcmin, $\sim 5$ kpc) which 
is also disturbed by the interaction of the jet and the inner 
radio lobes, where a narrow range of temperatures 
(well within $0.8 - 2$ keV) is indicated 
in the spectra. One should note, that the results derived by
Matsushita et al. (2001) have in contrast to most other work been
obtained from deprojected spectra in concentric rings, which 
feature within the validity of the assumption of approximate
spherical symmetry the spectral emission of three-dimensional 
volume elements. In fact the practically perfect fit with isothermal
spectra was only achieved after deprojection. The effort of
deprojecting the spectra is of course especially rewarding in the case
of M87 due to the high spatial resolution (8 arcsec correspond to
about 0.7 kpc at M87) and the excellent photon statistics
(for e.g. the PN more than 12 Million photons were received in
the useful observing period). 

It was also discovered in the analysis of the M87 X-ray spectra
that the emission from the innermost region ($r \le 1$ arcmin)
is most probably effected by resonant line scattering in some of
the most prominent emission lines, an effect that severely disturbs
the measurement of element abundances (B\"ohringer et al. 2001).
This could in principle be made responsible for the lack of some of
the spectral lines indicative of low temperature material in the 
cooling flow as observed with the RGS instrument. 
In the detailed analysis of the EPN and EMOS spectra
this problem could be avoided, however, by integrating over a large enough 
radial region that the net effect of resonant line scattering is 
negligible (almost all the scattered light will be reemitted close
to the center), and therefore it could be established that the 
lack of low temperature emission lines is not an artefact due to
resonant line scattering (Matsushita et al. 2001).  
                                                                                
\subsection{The Fe-L-shell lines as ICM thermometer}

\begin{figure}[h]

\psfig{figure=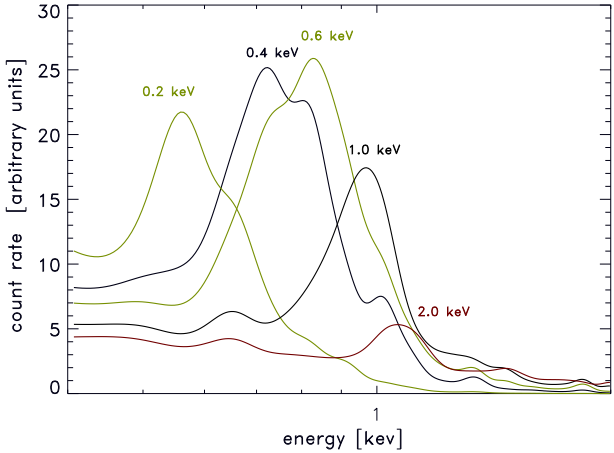,height=6.2cm}
\caption{The Fe L-line complex in X-ray spectra as a function of the plasma
temperature for a metallicity value of 0.7 solar. The simulations
show the appearance of the spectra as seen with the XMM EPN.
The emission measure was kept fixed when the temperature was varied.
}\label{fig1}
\end{figure}

Among the spectroscopic signatures which are sensitive to the 
temperature of the hot plasma in the temperature range relevant
to cooling flows, the complex of iron  L-shell lines is most
important. Therefore we will illustrate its use as a sensitive
thermometer of the intracluster temperature structure in more detail.
It has e.g. also been used in the {\sl XMM-Newton} study of the temperature
structure of the X-ray surface brightness excess features at the
location of the inner radio lobes of M87 by Belsole et al. (2001).
Fig.~\ref{fig1} shows simulated X-ray spectra 
as predicted for the XMM EPN instrument in the spectral
region around the Fe L-shell lines for a single-temperature plasma
at various temperatures from 0.4 to 2.0 keV and 0.7 solar metallicity
as measured for the iron abundance in the central region of the
M87 halo. There is a very obvious shift in the location
of the peak of the blend of iron L-shell lines. This energy change is caused
by the fact that with decreasing temperature the degree of ionization
of the Fe ions also decreases. The resulting effect is an increased
screening of the charge of the nucleus and a decreased binding energy
of the L-shell electrons which is reflected in a decreasing transition
energy for L-shell lines. This effect turns out to provide an excellent
thermometer in particular for the energy range shown. 

\begin{figure}
\psfig{figure=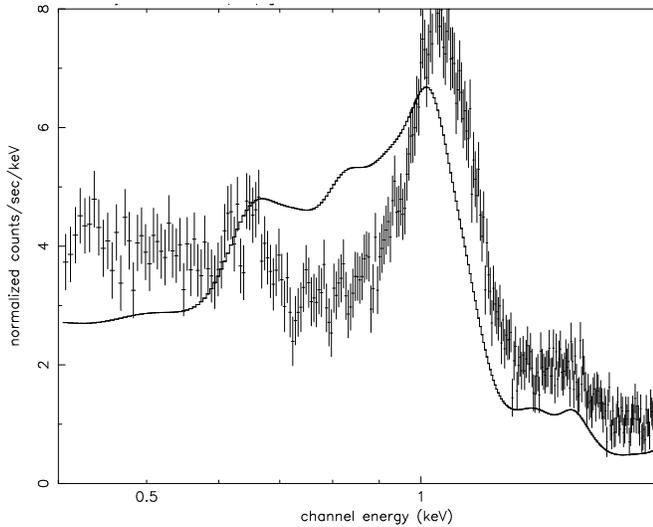,height=7cm}
\caption{{\sl XMM EPN}-spectrum of the central region of the M87 X-ray halo
in the radial range $R=1 - 2$ arcmin. The spectrum has been fitted 
with a cooling flow model with a best fitting mass deposition rate of
0.96 M$_{\odot}$ yr$^{-1}$ and a fixed absorption column density of
$1.8 \cdot 10^{20}$ cm$^{-2}$, the galactic value, and a parameter
for $T_{low}$ of 0.01 keV.}
\label{fig2}
\end{figure}

For a cooling flow with a broad range of temperatures one expects a composite
of several of the relatively narrow line blend features, resulting in 
a quite broad peak for the blend of iron L-shell lines. 
Fig. ~\ref{fig2} shows for example the deprojected 
spectrum of the M87 halo plasma for the radial range 1 - 2 arcmin
(outside the excess emission region at the inner radio lobes) and
a fit of a cooling flow model with a mass deposition rate slightly less
than 1 M$_{\odot}$ yr$^{-1}$ as approximately expected for this radial
range of the cooling core from the analysis of the surface
brightness profile (e.g. Stewart et al. 1984, Allen et al. 2001,
Matsushita et al. 2001). We have selected this radial region
to best demonstrate the diagnostics of the cooling core, since it
is far enough away from the central nucleus and jet not to
be contaminated by these non-thermal X-ray sources, still 
very centrally located in the cooling core, and providing a good 
photon statistics due to the high surface brightness.

\begin{figure}                                                                  
\psfig{figure=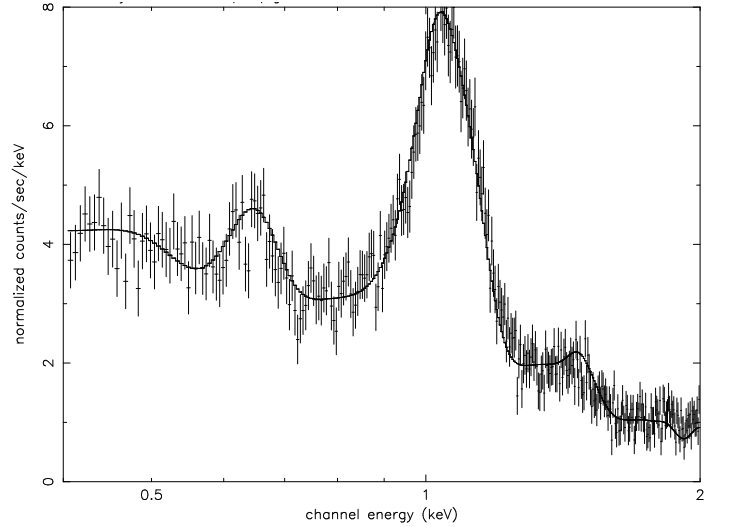,height=7cm}
\caption{{\sl XMM EPN}-spectrum of the central region of the M87 X-ray halo
in the radial range $R=1 - 2$ arcmin fitted by a cooling flow spectrum
artificially constraint to emission from the narrow 
temperature interval 1.44 - 2.0 keV yielding a mass deposition rate of
2.4 M$_{\odot}$ yr$^{-1}$. The parameter $T_{low}$ was treated as 
a free fitting parameter.}
\label{fig3}
\end{figure}

It is evident that the peak
in the cooling flow model is much broader than the observed spectral
feature. For comparison Fig.~\ref{fig3} shows the same spectrum fitted
by a cooling flow model where a temperature of 2 keV was chosen for
the maximum temperature and a suitable lower temperature cut-off was 
determined by the fit. As expected from the observed narrow peak shape,
only a narrow temperature range is allowed by the fit with a lower 
temperature cut-off at 1.44 keV. This is only meant as an example, and
the value for $T_{low}$ depends here slightly on the value chosen for the
fixed value for $T_{max}$. For a detailed discussion of the best 
constraints on isothermal model fits to the complete spectra
see Matsushita et al. (2001).
The discrepancy between the predictions of the cooling flow model
and the observations is indeed striking and even more obvious than
the discrepancies found for the RGS spectra of the more massive and 
hotter cooling flow clusters mentioned above. In the next two subsections
we therefore explore possibilities to reconcile the observations 
with the cooling flow expectations.

\subsection{Inhomogeneous metal abundances in the ICM}

One of the effects that could possibly rectify the standard cooling flow
scenario with the observed spectral properties of cooling flow regions,
as suggested by Fabian et al. (2001a), is a very inhomogeneous distribution
of the metal abundances in the cluster ICM. From the above discussion
it is obvious that the spectral diagnostics are essentially based
on the observation of spectral lines from metal elements. Therefore
without a significant metal abundance the signatures searched for are
absent. The model proposed by Fabian et al. is a genial 
recognition that the intensity of the metal lines in the
spectra is not simply proportional to the abundance. Since the total
amount of radiation power that can be emitted by the cooling gas
is limited by the gas's heat capacity, there is a saturation effect
in the line intensities, if the fraction of the emission power
contributed by the lines is large. This situation is most severe
at low temperatures. The application of this effect to cooling 
flow models where the total emission power for each temperature phase
is fixed by the mass deposition rate through eq.(1) is most 
interesting: the line intensity reduction due to an inhomogeneous
metal distribution is most effective for the low temperature phases,
an effect that could in principle be responsible for the non-detection
of the low-temperature lines.  

We can cast this argument into mathematical form to explore in 
more detail how the apparent abundance distribution is affected
by an inhomogeneous metallicity of the intracluster plasma.
In the following calculation we investigate in a simplified but
very illustrative model how the apparent metallicity changes 
if we start with a mean metallicity, $z$, which is distributed 
unevenly in two plasma phases with the 
same temperature. We make the further simplifying assumption that
phase 1 is completely devoid of metals and phase 2 is enriched. 
We calculate the apparent metallicity that is derived 
if the resulting combined spectrum is analysed 
under the assumption of a chemically homogeneous 
plasma. The apparent metallicity is proportional
to the observed intensity ratio of the lines to the continuum. The
change in the apparent metallicity is thus equivalent to the suppression
factor of the line signature of the cooling flow at low temperature.
This two phase model is an approximate and simple representation of 
the scenario described by 
Fabian et al. (2001a) in which the minor ICM component is heavily
enriched by recent SN Ia and stellar wind contributions while the bulk
of the gas has relatively low metallicity.

For the model we define the following important parameters: $z$ is the mean
metallicity in solar units as defined above, 
$f$ is the mass fraction of the enriched
plasma phase, $z_2 = z \cdot f^{-1}$ is the metallicity of the enriched
phase, and $<z_{obs}>$ is the mean apparent (``observed'') metallicity.
We further define the parameter $\cal F$ as the fraction of the emitted 
flux in metal lines as compared to the total emission, and  ${\cal F}_1$
as this fraction for the case of solar metallicity ($z = 1$) of a 
homogeneous medium. ${\cal F}_2$ then designates the line intensity fraction
for the enriched second phase of the inhomogeneous ICM model.

\begin{figure}                                                                  
\psfig{figure=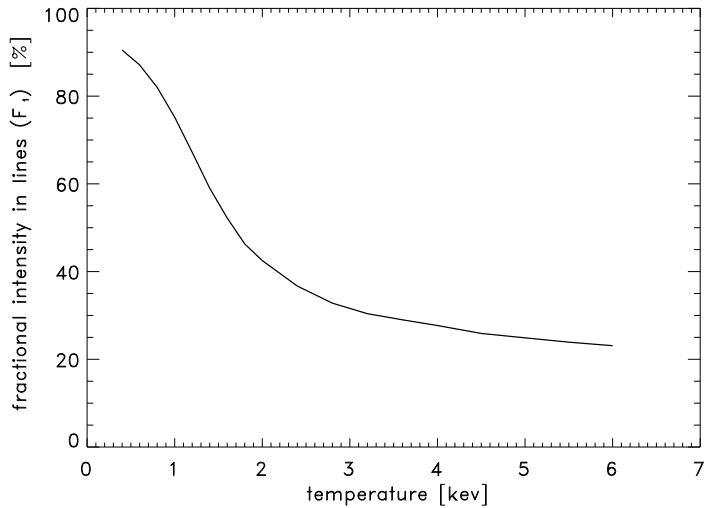,height=7cm}
\caption{Fraction of radiative emission power that is contributed by
the metal ions in the plasma, ${\cal F}_1$, as a function  of temperature for
a solar abundance plasma (with abundances from Anders \& Grevesse 1989).}
\label{fig4}
\end{figure}

The parameter ${\cal F}_1$ is an important indicator that divides the
metallicity range in which  $\cal F$ is proportional to $z$ and
for which  $\cal F$ is in the saturation regime. The values of ${\cal F}_1$ 
as a function of plasma temperature in the temperature
range relevant for the present discussion are given in Fig.~\ref{fig4}. 
The formulae used to calculate $\cal F$ as well as the basics of the
following calculations are derived in detail in the Appendix.
We note in Fig.~\ref{fig4} that ${\cal F}_1$ is increasing with decreasing
temperature, the saturation effects leading finally to a suppression
of the apparent, observed metallicity will therefore be most effective
at low temperatures as suggested by Fabian et al. (2001a).

\begin{figure}                                                                  
\psfig{figure=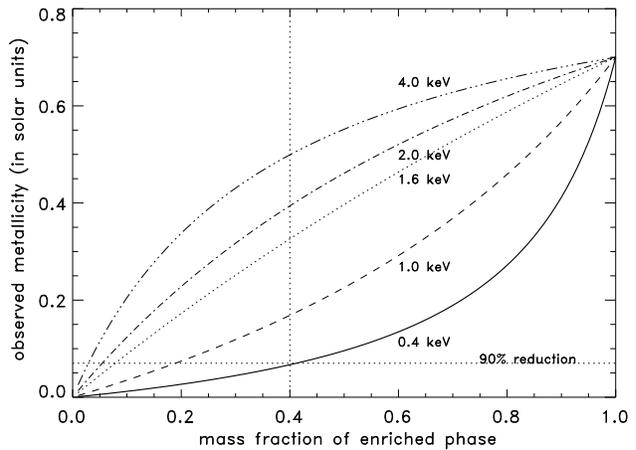,height=6.2cm}
\caption{Apparent deduced metallicity from the spectrum emitted
from a chemically inhomogeneous two-component plasma as a function
of temperature when a single-phase plasma is assumed in the analysis.
The mean metallicity in the model is 0.7 solar. Results for different 
plasma temperatures are given in the plot. For further explanations
see the text.}
\label{fig5}
\end{figure}

The apparent, observed metallicity can then be calculated by means
of the following formula:

\begin{equation}
<z_{obs}> = {z~(1 -{\cal F}_1) \over 
1~+~ z{\cal F}_1~(z f^{-1} - z - 1)}
\end{equation}

Fig.~\ref{fig5} shows the resulting apparent metallicity as a function 
of the mass fraction (equal to the emission measure fraction) of the
enriched gas. Here the mean metallicity, $z$, has been taken
to be 0.7 of the solar value as measured for iron in the central region
of the M87 X-ray plasma halo. We note indeed, that the suppression of 
the metal lines is very dramatic for the low temperatures. Therefore
this effect should in general be taken very seriously and for each case
of a metallicity determination at these low temperatures we have to
check carefully if a homogeneous chemical distribution can be assumed
or if corrections for an inhomogeneous distribution might apply. 

\begin{figure}
\psfig{figure=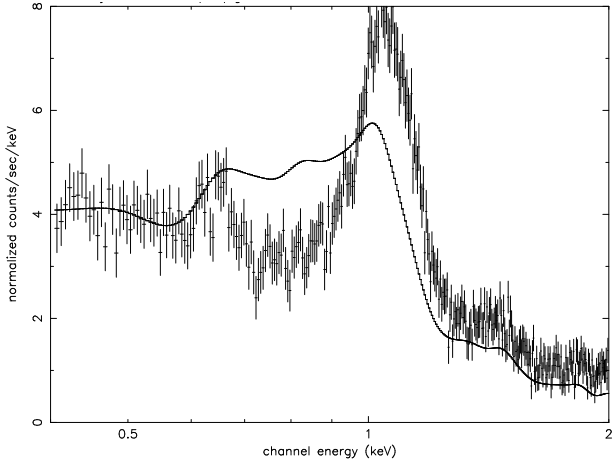,height=7cm}
\caption{{\sl XMM}-spectrum of the 1-2 arcmin region in the M87 halo
fitted by a model with two cooling flow phases, one with very large
metallicity (about 5 times the solar value) and one with no metals.
The normalization of the two phases is roughly equal and the upper
temperature limit is 2 keV.} 
\label{fig6}
\end{figure}

Inspecting Fig.~\ref{fig5} we note two interesting effects. Firstly,
if we want to suppress the line emission e.g. for the temperature phase 
with $T = 0.4$ keV by an order of magnitude as would approximately 
be required to bring the cooling flow prediction for this 
temperature phase in accord with the observed spectrum for M87, 
the apparent metallicity for the bulk temperature of 1.1 to 1.6 keV
in the inner cooling flow region would also be a factor of 2 to 4
lower than the true metallicity. The metallicity increase towards
the center would than be even more dramatic than already derived 
for a homogeneous metallicity distribution with an increase of the
iron abundance by a factor of 2 to 3. This can be seen e.g. in 
Fig. 6 where the fit yields a metallicity of 5 times the solar value 
for the enriched phase which has roughly half the emission measure.
Secondly the suppression of the
metal lines is an effect which gradually increases with decreasing
temperature, so that if strong Fe L-shell lines are observed at  
1.1 to 1.6 keV they will not suddenly disappear from the spectrum at 
slightly lower temperature. Thus we would still expect to observe a 
considerably broadened iron L-shell peak as was shown in Matsushita
et al. (2001). We show this effect also here in Fig.~\ref{fig6} 
for the example
of the X-ray emission from the $r= 1' - 2'$ ring in M87. The spectral
iron feature is clearly not consistent with the observed spectrum.
Since we consider these examples as representative of the general 
solution of the problem with a variety of possible mixing distributions
of the metal abundances, we do not expect that the general  discrepancy 
can be resolved by adopting a chemically inhomogeneous intracluster
medium (at least for this detailed observation of M87).

\section{The Interpretation of Cooling Flow Spectra and Cluster Intrinsic
Absorption}

\begin{figure}
\psfig{figure=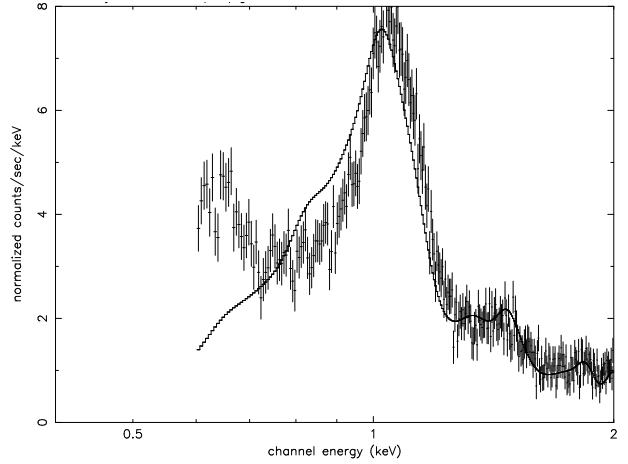,height=7cm}
\caption{{\sl XMM EPN}-spectrum of the central region of the 
M87 halo as shown in Fig.~\ref{fig2}, now fitted with 
a cooling flow model and a free parameter
for the internal excess absorption. The spectrum was constrained to
the temperature interval 0.6 to 2.0 keV.}
\label{fig7}
\end{figure}

Another possible way to make the observed Fe-L line feature roughly
consistent with a cooling flow model is to allow the absorption 
parameter in the fit to adjust freely. This is demonstrated in 
Fig.~\ref{fig7} where we have again taken our example spectrum 
of the annular region in the M87 halo. Now we have limited the
spectral range to be fitted to energies above 0.6 keV and taken
the absorption column density as a free fit parameter. In order to
match the Fe-L peak shape better, the best fitting absorption column 
density is selected in such a way by the fit that the absorption edge
limits the extent of the Fe-L line feature towards lower energies.     

This is actually a good representation of the spectral fitting of cooling flow 
models that have been performed on {\sl ASCA} data. For {\sl ASCA} the
sensitivity for the low energy part of the spectrum was much less than
for {\sl XMM-Newton} and this spectral energy range was not very well
calibrated. Therefore, in general the spectral part below about 
0.6 to 0.7 keV was not included in the fit (e.g. 
Allen 2000, Allen et al. 2001), just as done in our example. Therefore
the {\sl ASCA} experience has left us with two possible options of the
interpretation of the spectra of cooling core regions: (1) an interpretation
of the results in form of an inhomogeneous cooling flow model which
than necessarily includes an internal absorption component
(e.g. Fabian et al. 1997, Allen 2000, Allen et al. 2001), or (2)
an explanation of the spectra in terms of a two-temperature component
model (e.g. Ikebe et al. 1997, 1999, Makishima 2001) where the hot 
component is roughly equivalent to the hot bulk temperature of the
clusters and the cool component corresponds approximately to 
$T_{low}$. As actually pointed out e.g. in Ikebe et al. (1999) and also seen
in Allen et al. (2001) the two-temperature models provided in almost
all cases a slightly but significantly better fit to the spectra
as measured by the $\chi ^2$ statistic. 

\begin{figure}
\psfig{figure=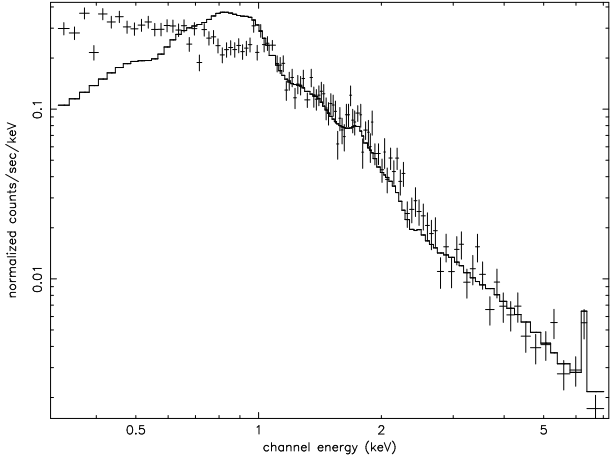,height=7cm}
\caption{{\sl XMM EPN}-spectrum of the central region of the cluster A1795
fitted by a cooling flow model with a mass deposition rate of 
50 M$_{\odot}$ yr$^{-1}$ and an additional hot component at a
temperature of 5.2 keV. The value for the absorbing column density
was set to the galactic value of $1.26\cdot 10^{20}$ cm$^{-2}$. }
\label{fig8}
\end{figure}

\begin{figure}
\psfig{figure=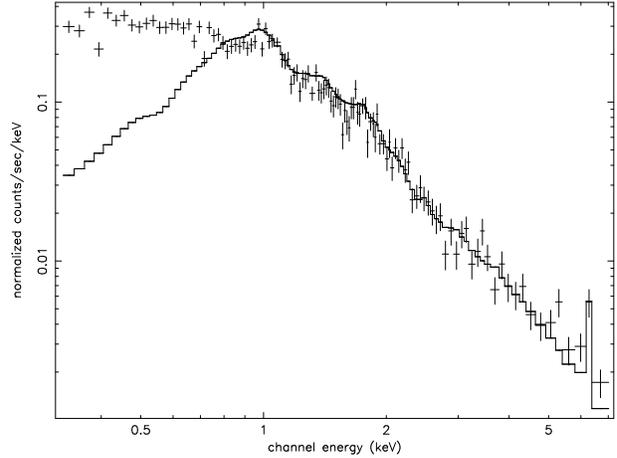,height=7cm}
\caption{{\sl XMM EPN}-spectrum of the central region of the cluster A1795
fitted by a cooling flow model with a mass deposition rate of
50 M$_{\odot}$ yr$^{-1}$ and an additional hot component at a
temperature of 5.2 keV. The value for the absorbing column density
of $3 \cdot 10^{21}$ cm$^{-2}$ was chosen to obtain an approximate
fit to the narrow iron L-shell line feature.}
\label{fig9}
\end{figure}

This phenomenon is not limited to the case of M87, even though this
is by far the best case where things can be demonstrated in the
most obvious way, mostly due to the good (absolute) spatial resolution 
and photon statistics of this observation. In Figs.~\ref{fig8} and 
\ref{fig9} this effect is illustrated for the cooling core region of
the hotter and much more massive cluster A1795 which also harbours
a much stronger cooling flow (e.g. Perez et al. 1998). Fig.~\ref{fig8}
shows the fit of a standard cooling flow model (with a very moderate
cooling flow of only 50 M$_{\odot}$ yr$^{-1}$ compared to the
value of $ \sim 340 - 750$  M$_{\odot}$ yr$^{-1}$ 
deduced from the imaging data by Peres et al. 1998) without absorption.
Again we note a broad Fe-L line feature from the cooling flow model
which is not observed. This feature
is constraint to a narrower peak in Fig.\ref{fig9} due to the inclusion
of an internal absorption component of the order of $3 \cdot 10^{21}$
cm$^{-2}$.

Thus there is a combination of the effects of internal
absorption and the signature of cooling flows which can be 
interpreted as a suspicious coincidence or as a reconfirmation of 
the cooling flow model due to the detection of the cold material sink.
We have seen in the above examples that 
in order to produce a sharp Fe-L line feature the absorption edge
has to appear at the right energy and therefore values for the 
absorption column of typically around $3 \cdot 10^{21}$ cm$^{-2}$
are needed (for nearby clusters). 
Values of this order are actually found in most
of the cooling flow model fits. For example Allen (2000) and 
Allen et al. (2001) find in recent improved {\sl ASCA} studies  
indeed values in the range $1.5 - 5 \cdot 10^{21}$ cm$^{-2}$. 

Contrary to the case of {\sl ASCA},  the {\sl XMM-Newton} data can be 
analysed to lower energies and the presence of the strong soft
continuum component (including the oxygen Ly$\alpha$ line in
the case of M87) indicates that absorption may not be present
or is affecting only the cold and dense part of the cooling core
plasma. Therefore it is very important to test the case of 
internal excess absorption in an independent way. This is now made
possible for the first cases thanks to the {\sl XMM-Newton} and
{\sl Chandra}  observatories as explained in the next section.
          
\section{Non-Detection of Sufficient Internal Absorption in Two Test Cases}

\begin{figure}
\psfig{figure=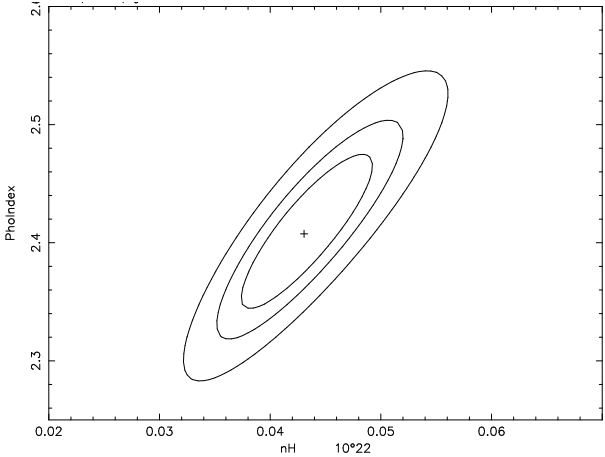,height=6cm}
\caption{Constraints on the shape of {\sl XMM EPN}-spectrum of 
the nucleus of M87. The lines show the 1, 2, and 3$\sigma$ confidence
intervals for the combined fit of the slope (photon index) of the
power law spectrum and the value for the absorbing column density,
$n_H$.}
\label{fig10}
\end{figure}

\begin{figure}
\psfig{figure=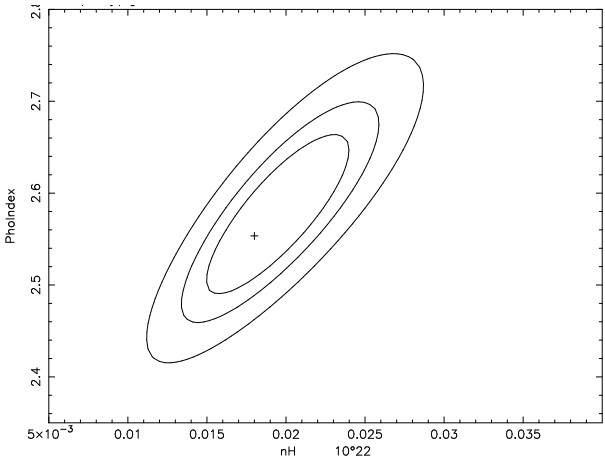,height=6cm}
\caption{Constraints on the shape of {\sl XMM EPN}-spectrum of 
the brightest knot in the jet of M87. 
The lines show the 1, 2, and 3$\sigma$ confidence
intervals for the combined fit of the slope (photon index) of the
power law spectrum and the value for the absorbing column density,
$n_H$.}
\label{fig11}
\end{figure}

To perform a test on the presence of absorbing material in the central regions
of clusters independent if the complex fitting of multi-phase
intracluster medium emission models, we need an independent light source.
This fortunately exists in the form of the emission from the central
AGN in many cooling flow clusters (e.g. Burns 1990). 
For example in the case
of M87 we can now obtain spatially resolved spectra of the nucleus
and the brightest knot in the jet with the high collecting power
of {\sl XMM-Newton} (B\"ohringer et al. 2001). These spectra can
quite well be fitted with a power law spectrum and nearly galactic 
absorption. In Figs.~\ref{fig10} and \ref{fig11} we show the combined
constraints on the power law slope and the absorbing column density
from the spectral fits. The $2\sigma$ upper limit on the excess absorption
is with a value of about $5 \cdot 10^{20}$ cm$^{-2}$ almost one 
order of magnitude
lower than what is required for a successful cooling flow spectrum fit.

\begin{figure}
\psfig{figure=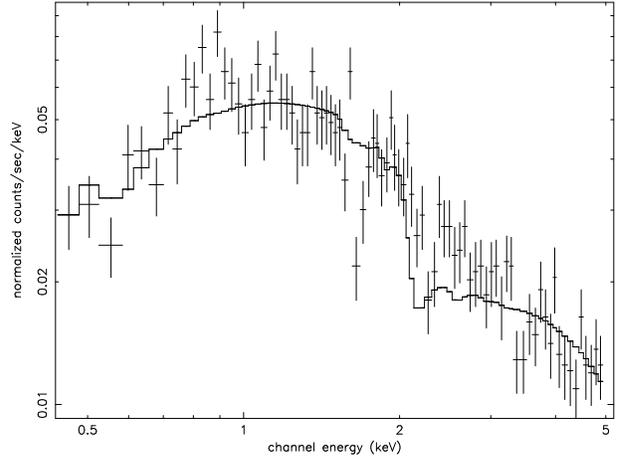,height=7cm}
\caption{{\sl Chandra}-spectrum of the nucleus of NGC1275 fit by a 
power law spectrum and a free fitting parameter for the absorbing column,
for which a best fitting value of about $6\cdot 10^{20}$ cm$^{-2}$ was
found.}
\label{fig12}
\end{figure}

\begin{figure}
\psfig{figure=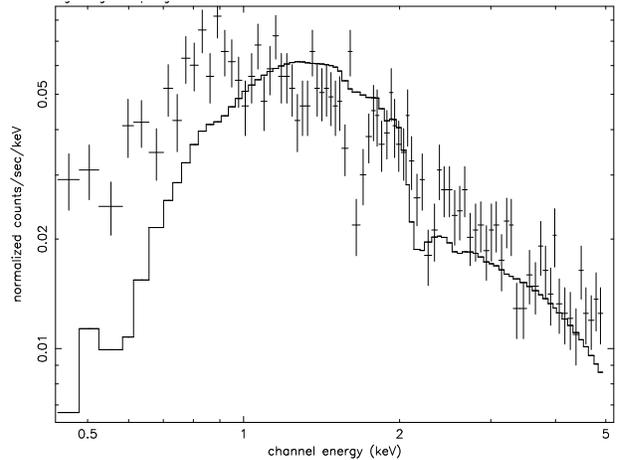,height=7cm}
\caption{{\sl Chandra}-spectrum of the nucleus of NGC1275 fit by a 
power law spectrum and a fixed fitting parameter for the absorbing column
density of $3.3 \cdot 10^{21}$ cm$^{-2}$ as approximately needed in the fits
of cooling flow models to {\sl ASCA} spectra of NGC1275.}
\label{fig13}
\end{figure}

Similarly the central AGN in NGC1275 in the center of the Perseus
cluster is resolved with {\sl Chandra} and we can obtain a spectrum
from this point source (from the Perseus observation data in the     
{\sl Chandra} archive) as shown in Figs.~\ref{fig12} and \ref{fig13}. 
These spectra are much noisier than those for M87 due to the lower
photon statistics. The best fitting value for the absorbing
column density is even slightly lower than the galactic value.
An absorption column of more than $3 \cdot 10^{21}$ cm$^{-2}$ as needed
for a successful fit of the cooling flow model is clearly inconsistent
with the observations.
There is some concern that the spectrum of the NGC1275 nucleus
is affected by photon pile up in the CCD pixels. The count rate
for the nuclear emission is with a value of about 0.07 counts s$^{-1}$
quite low, however, and according to the {\sl Chandra} Handbook
the effect is in the percent range which cannot explain the large
difference between the high absorption fit and the observed spectrum.
This is all the more true as the effect is expected to result in
low energy photons being sometimes bunched together and registered
as events of higher energy, thus reducing the low energy flux 
mimicking a higher absorption, which is not found in the observations.   

Hopefully this test can be extended to further cooling core regions
by means of {\sl XMM-Newton} and {\sl Chandra} to see if this conclusion
can be reached in general. Nevertheless these two test cases indicate
that the agreement of the cooling core spectra with the inhomogeneous
cooling flow model plus excess absorption may have been obtained through
a fortunate conspiracy of the two effects and the new results 
(including the RGS, EPN, and MOS spectra for a number of clusters)
are no longer supporting this interpretation. In one further case,
in Hydra A, the spectrum of the central nucleus has been resolved
and recorded. Unfortunately in this case a very high absorption value
in excess of $10^{22}$ cm$^{-2}$ has been found (McNamara et al. 2000)
which has to originate from very close to the nucleus and is not caused
by the cooling core region. But due to the high AGN intrinsic absorption
it does not allow us to probe the possible lower absorption of the 
cooling flow.  

There is still another argument left to save the absorption picture,
which is a partial covering model, where the absorption would only
be effective in the line-of-sight of the cold clumps. Even though
this might be considered as an unwanted anesthetic complication
of the cooling flow model it should nevertheless be considered.
Thus lets assume that the absorbing material is only found in
connection with cold condensing gas and lets again scale the model
to the M87 halo where we have the best observable parameters.
If most of the hotter, visible plasma is confined to the temperature 
range 0.8 to 1.7 keV and the hidden plasma to temperatures below the
lower limit of this range, the colder plasma will be approximately
a factor of 2 denser and will cool about a factor of 4 faster
(due to the higher density and the higher efficiency of the line cooling).
The cool gas will than occupy a volume fraction of the order of 10\%.
Lets further assume that the enriched phase constitutes 
only about 40\% of this cold plasma (a lower value would in the light of
the above results lead to very high ``true'' metal abundances and
thus we consider 40\% as an extreme limit). Thus the material
associated to the intrinsic absorption effect should at least
have a filling factor of about 4\%. Already the deep pointings
on M87 with the ROSAT HRI should have resolved cool clumps with
kpc size in the surface brightness distribution of the M87 halo
image, thus these condensations must have a smaller scale. Taking
the total line-of-sight through the cooling flow region starting
at a cooling radius of about 50 kpc, we find that the optical depth
for having clumps of kpc size in the line-of-sight towards the nucleus is about 2.
Thus the chance that we have missed any absorbing column in the
three tests with the nuclei of M87 and NGC 1275 and the jet of M87
is indeed small. Nevertheless, it will be very important to find 
further test cases. 
  
\section{Evidence for Inhomogeneities and Mass Deposition in X-ray Images}

In view of the problems encountered above in the interpretation of 
the recent observational data in terms of a standard cooling flow
model, we will explore in this section and the next one two more
general aspects of the cooling flow model which hopefully will
give further guidance to the interpretation of the observational 
results of cluster cooling cores:
(i) observations of inhomogeneities in the inhomogeneous 
cooling flows and (ii) cooling flows in rotating clusters.

The inhomogeneous cooling flow model features a thermally unstable
plasma in approximate pressure equilibrium where the cooling 
of the gas is accelerating when the gas is cooling. That is,
temperature and density differences are magnified during the 
cooling process. Since initial density inhomogeneities are required
to obtain an inhomogeneous cooling flow and since these inhomogeneities
are continuously increasing in the comoving cooling flow model 
without heat exchange, we should expect that the cooling flow zone
is characterized by a very inhomogeneous plasma. Inspecting some
of the best observed cooling flow clusters, e.g. M87, we find that
the cooling flow regions actually have a quite regular X-ray appearance
over most of the cooling flow region and in most cases they have  
elliptical symmetry reproducing the potential of the underlaying cD
galaxy almost perfectly (e.g. Allen et al. 1995, B\"ohringer et al. 1997,
Buote \& Canizares 1994, 1996). We have sometimes argued in the past that
this just reflects the fact that the inhomogeneities in the cooling flow
have to appear on scales not yet resolved by the available X-ray imaging
instruments and used this to set a lower limit on the suppression
of classical heat conduction as required by the cooling flow model
(e.g. B\"ohringer \& Fabian 1989). 
Any detection of such inhomogeneous structure would help
to guide us in the more realistic modeling of cooling flows. 

\begin{figure}                                                                  
\psfig{figure=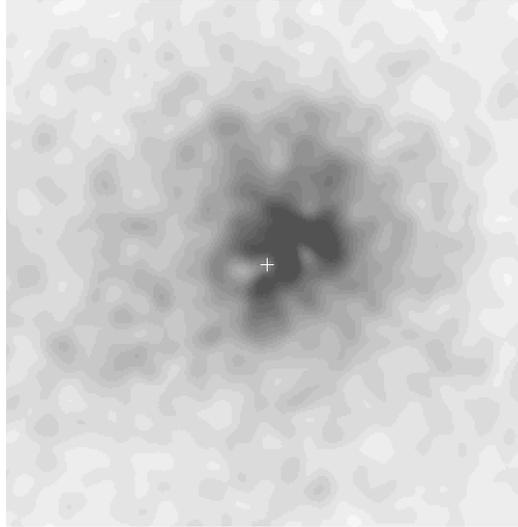,height=7cm}
\caption{X-ray morphology of the central part of the M87 halo
as observed with the {\sl ROSAT HRI}. The two point sources 
originating from the central AGN and the brightest knot of the 
jet have been subtracted. The position of the nucleus of M87 is 
marked.  The scale of the image is $2.13 \times 2.13$ arcmin$^2$. 
The X-ray emission is clearly not symmetric around the 
galactic center of M87 and there is an obvious emission excess
towards the North of the nucleus where also emission 
line filaments are observed.}
\label{fig14}
\end{figure}

A deeper inspection of some of the closest clusters observed with
the {\sl ROSAT HRI} and now with high resolution images taken with
{\sl Chandra} actually reveals inhomogeneous structure in the 
cluster cores. In Fig.~\ref{fig14} we show for example the 
{\sl ROSAT HRI} image of the core region of the M87 X-ray halo.
The point sources originating from the M87 AGN and the brightest knot
in the jet have been removed from the image. One clearly notes that
the X-ray emission is not symmetric around the AGN, which is most
probably the gravitational center of the cD galaxy M87. The surface
brightness distribution can be interpreted as a region
of diffuse excess emission in the North of the nucleus (B\"ohringer
1999). 

A quite similar surface brightness excess region is found in the 
{\sl ROSAT HRI} and {\sl Chandra} images of NGC1275 towards the SE
of the nucleus of this galaxy (B\"ohringer et al. 1993, 
Fabian et al. 2000). 
In the case of M87 the analysis of
the X-ray hardness distribution in the {\sl XMM-Newton} data indicates 
that this excess region is colder than the surroundings 
(Matsushita et al., in preparation) and for the bright blob in NGC1275
the {\sl Chandra} data indicate the same (Schmidt \& Fabian 2001).
Therefore these excess emission regions most probably are colder
and denser blobs. They could be inhomogeneities in which material is 
presently cooling and condensing in the cooling cores of these clusters.
Further evidence for such inhomogeneities is now coming from 
many high resolution images of cluster cooling cores of nearby 
clusters which can be well enough resolved
with {\sl Chandra}. In A1795 for example
a cooling tail has been found with a length of about $80 h_{50}^{-1}$ kpc
and temperatures in the range 2 - 3 keV, cooler than the ambient medium 
(Fabian et al. 2001b). The X-ray filament coincides with an emission
line filament observed in H$\alpha$+N$II$ (Cowie et al. 1983).
In Hydra~A McNamara et al. (2000) observed a flattened excess X-ray emission
structure which is coincident with a gaseous star-forming disk 
with an estimated star formation rate between 1 and 15 M$_{\odot}$ yr$^{-1}$
(depending on the duration of the assumed star burst $\sim 10^9$ or $10^8$
yr, respectively, McNamara et al. 2000). Also in A1835 an excess emission 
region close to the center and inhomogeneities are observed. These features
are well inside the optical central galaxy of the cluster (Schmidt et al. 2001).

The common feature of almost all these detections of inhomogeneities is that
these features are found in the very central region of the cooling flow,
most often inside the optical image of the central cD galaxy. The filament
in A1795 with a length of $80 h_{50}^{-1}$ kpc is among these examples 
by far the largest off-center structure. Thus we note that these 
inhomogeneities are generally found only in the centers of cooling flows 
and most of the much larger cooling flow region (with radii from 50 kpc
to over 200 kpc for the above examples, see Peres et al. 1998, 
Schmidt et al. 2001) is generally more regular.

Taking these features as a manifestation of actually cooling and condensing gas
we find that the mass deposition rates calculated for
these localized regions is (of course) much smaller than what is inferred for
the entire cooling flow.  In the case of the bright blob north of the
center of M87 we find a steady state mass deposition rate in the 
blob if it is cooling without disturbance of a few tenths of a solar mass
per year, while for the bright blob in NGC1275 the estimated mass deposition
rate is of the order of 10 solar masses per year. Thus, mass deposition,
where we can supposedly observe it directly, is one or two orders
of magnitude less than what has classically been claimed for the cooling 
flow mass deposition. If steady
state cooling would be reduced to these low values, there would be no
conflict with the interpretation of the spectroscopic results since
in most cases a cooling process is consistent with the data if the
mass deposition rate is reduced by at least one order of magnitude
with respect to the conventional cooling flow value (see e.g. Matsushita
et al. 2001 for the case of M87 where a mass deposition rate of up
to 1 M$_{\odot}$ would be consistent with the data).

Lower mass deposition rates in cooling flows have been advocated
previously by a number of authors e.g. O'Dea (1994), Braine et al. (1995),
Voit \& Donahue (1995), McNamara et al. (2000), 
mostly to bring the
results into consistency with the observations of optical emission lines
and indications of star formation. Even the very recent observations
of large amounts of warmed molecular gas are not necessary calling
for very high mass deposition rates. For 34 clusters observed by
Edge (2001) only 7 extreme cooling flow clusters show mass estimates
for the cold gas from CO line observations in excess of 
$3 \cdot 10^{10}$ M$_{\odot}$ most of
which have inferred mass deposition rates of 1000 M$_{\odot}$ yr$^{-1}$
and more. Thus, if the gas observed in CO is the final cold sink
it could still be explained by mass deposition rates reduced by
an order of magnitude and for the smaller inferred gas mass values
stellar mass loss could even be an important contribution.  

\section{Problems of Cooling Flows in Rotating Clusters}

The effect of cluster rotation (or the rotation of elliptical galaxies)
on cooling flows has been considered in several works 
(e.g Kley \& Mathews 1995, Garasi et al. 1998). Since
we have not yet found much direct evidence for cluster rotation, the
implications of these studies have not generally been integrated into the 
cooling flow picture. We approach this problem here from the general
perspective of structure formation in the Universe from which statistical
predictions for the degree of rotation to be expected for clusters are
obtained and then discuss the implications for massive cooling flows.

Simulations of the evolution of structure provide predictions for the 
typical spin parameter, $\lambda$, of dark matter halos, e.g. in the 
scale-free simulations of Efstathiou et al. (1988) with values of about
$\lambda = 0.04 -0.06$ (where the spin parameter is defined as 
$\lambda = J~ |E|^{1/2}~ G^{-1}~ M^{-5/2}$; see also e.g. White 1994
for an overview). Efstathiou et al. (1988) also provide the more
practical values of the ratio of the rotational velocity to the velocity
dispersion e.g. for the region where the overdensity is about 500
of the critical density of the Universe, with a value of about 0.15.
In galaxy clusters this rotation has not been observed directly but
it has been investigated in detail for giant elliptical galaxies
which constitute smaller dark matter halos which can be viewed as
scaled down versions of clusters.
In a study of 22 ellipticals with a mean velocity dispersion
of about 250 km s$^{-1}$ by Franx et al. (1989) they find a 
mean rotational velocity of
about 52 km s$^{-1}$ which is slightly larger but 
roughly consistent with the value of
about 15\% found in the simulations. The slightly higher value 
actually provides some margin to include some energy dissipation 
and matter segregation in this case.        

In the case of M87 there is actually some weak indication for 
large-scale rotation
of the dark matter halo from the velocity distribution
of the globular cluster system associated to this giant elliptical. 
Some indication
for rotation was found by Sembach \& Tonry (1996) and Mould et al. (1990).
In a recent study of the velocity distribution of 205 globular clusters
in M87 these earlier indications were confirmed and 
improved by Cohen \& Ryzhov (1997).
Even so, the scatter in the data is large, the best fitting model 
gives a significant rotation with  a rotational velocity
of the order of 100 km s$^{-1}$ roughly in the direction of the 
PA 140 $(\pm 30)\deg$ for a mean radius of about 15 kpc. This would,
compared to the velocity dispersion in the larger dark matter halo
of the Virgo cluster  
(in early type galaxies) with a value of about 550 to 600
km s$^{-1}$, be again consistent with the ratio of rotational to random
velocities in the inner region of the Virgo cluster.

To see how this affects the structure of cooling flows, let us
start with a model for M87 and then draw general conclusions.
In M87 the cooling radius for a cooling time of 10 Gyr is
at about 55 kpc. Thus we take a radius of about 50 kpc to start
the cooling flow. For this radius we calculate a Keplerian velocity 
of 720 km s$^{-1}$ and the observed velocity dispersion for early type
galaxies which trace the cluster potential best has the value quoted above.
With the ratio of the rotational velocity to the velocity 
dispersion of about 0.15 we find a typical rotational velocity to be expected 
of about 82 km s$^{-1}$. If the cooling flow starts with this rotation
and flows inward under conservation of angular momentum, the 
flow will be rotationally supported when it reaches a radius of
about 11 kpc when the Keplerian velocity estimated from the 
mass profile derived by Matsushita et al. (2001) is about 
375 km s$^{-1}$. At this radius we expect an accretion disk to
be formed. Due to the angular momentum that will be transported
outwards it should also extend to slightly larger radii. Thus in M87 
the accretion disk should at least have a radius of about 12 kpc
(2.4 arcmin). Inspecting Fig.~\ref{fig14}, there is no signature of a
rotationally supported gas disk of this size. 
If we take the rotation of the 
globular cluster system as a guide we should expect the disk to be
oriented roughly along PA 140 $\deg$. 

The general case for any cluster will be similar to the case of M87
where the cooling radius of about 50 kpc is probably also close
to the dark matter core radius. In general one finds cooling radii
in the range of about 0.5 to 1 times the estimated gravitational 
core radius. At the core radius the Keplerian velocity is very
close to the velocity dispersion of the system (see e.g. King type
models described in Binney \& Tremaine 1997). Starting with a 
rotational velocity of 15\% of the Keplerian velocity at the cooling 
radius and with the spinning up of the flow with decreasing radius, 
Keplerian velocity will be reached with less than a fivefold 
decrease in radius. Note that the Keplerian velocity decreases 
with decreasing radius inside the core radius of the dark matter potential.
For cooling radii in the range of 50 - 150 kpc as found for massive
cooling flows this should generally result in the formation of 
rotationally supported disks with sizes of at least 10 to 30 kpc.

In the simulation studies of cooling flows in rotating systems
by Kley \& Mathews (1995) and Garasi et al. (1998) flattened 
disks of hot gas were found, as expected. In both simulations the
X-ray appearance was explicitly calculated and very flattened
X-ray surface brightness distributions were predicted. 
These disks should be observable in the nearby clusters with high 
resolution X-ray imaging devices, at least in cases where
they are seen edge on. Those features have not yet been observed
in general, however, and we conclude that rotationally supported
disks are not a general phenomenon in cooling flows as would
be expected. 

Indications of some trace of rotation has possibly been 
observed in the center of the Perseus cluster, were a spiral
feature in the X-ray surface brightness distribution 
(Churazov et al. 2000) may indicate that we are looking at
a rotating structure face on (see also Fabian et al. 2000).
Also in Hydra A a small rotational structure of emission line gas
has been observed (e.g. Heckman et al. 1985). But with the exception
of Pereus perhaps, large rotational structures are not observed in general.

\section{Heating Model}

Most of the problems in the interpretation of the observational 
results in terms of classical cooling flow models mentioned above, e.g.
(i) the inconsistencies of the spectral models with the 
observed spectra, (ii) the lack of signatures of massive mass deposition
at the degree predicted by the cooling flow model from image analysis
at other wavelengths than X-rays and the lack of strong enough
star formation indices, (iii) the problem of angular
momentum in massive flows, can be avoided if the flow and the mass
deposition is reduced to a much smaller level than previously thought.
The reduction should at least be about one to two orders of magnitude.

To reduce the mass condensation in the presence of energy conservation
some form of heating is clearly necessary to balance the radiation
losses. In principle three forms of heating sources have been discussed
in the literature:

\begin{itemize}

\item heating by the energy output of the central AGN (e.g.
Pedlar et al. 1990, Tabor \& Binney 1993, McNamara et al. 2000,
Soker et al. 2001)

\item heating by heat conduction from the hotter gas 
outside the cooling flow (e.g. Tucker \& Rosner 1983, Bertschinger \&
Meiksin 1986)

\item heating by magnetic fields, basically through some 
form of reconnection (e.g. Soker \& Sarazin 1990, Makishima et al. 2001)
where in the model of Makishima et al. the energy dissipated by the
magnetic fields originates from the motion of the cluster galaxies.

\end{itemize}

The latter two processes depend on poorly known plasma physical
conditions in the intracluster medium of the clusters and therefore
most probably remain in their quantitative effect highly speculative. 
The energy
output of the central AGN can be determined, however, and we can 
obtain very clear boundary conditions for our model. Therefore
we concentrate in this paper on the exploration of central cluster AGN as a 
source for heating the cooling cores of clusters.  

Before starting the discussion of possible models we can put forward
some general considerations on the properties that a successful 
heating model should have to provide a consistent interpretation.

\begin{itemize} 

\item the energy input has to provide sufficient heating
to balance the cooling flow losses, that is about $10^{60}$ to 
$10^{61}$ erg in 10 Gyr or on average about $3 \cdot 10^{43} -
3\cdot 10^{44}$ erg s$^{-1}$. 

\item the energy input has to be fine-tuned. Too much heating 
would result in an outflow from the central region and the central
regions would be less dense than observed. Too little heat will
not reduce the cooling flow by a large factor. Therefore the 
heating process has to be self-regulated: mass deposition triggers
the heating process and the heating process reduces the mass 
deposition.

\item the energy deposition has to provide a global heating
effect. Local energy deposition would result in local heating
while the mass deposition can still go on in the less well heated
regions.

\item the heating process has to preserve the observed entropy 
structure in the intracluster medium in the central regions of
cooling core clusters, with an entropy profile decreasing 
towards the center. This rules out for example any model in which
heat is transported to larger radii by classical convective transport
driven by a decreasing entropy with radius.

\item The heating process should preserve the observed strongly
increasing metallicity gradients in cooling core clusters.

\end{itemize}

We will discuss briefly in the following how these five requirements can
be met. In a paper in preparation by Churazov et al (2001b) this model
will be worked out and discussed in more detail.
For the modeling of the heating process we take our guidance from 
the detailed observations of the interaction process of 
the AGN radio lobes in M87
and NGC 1275 with the intracluster medium of the Virgo and Perseus
cluster, respectively, and derive important boundary parameters
from these observations. The interaction model is based on the 
scenario of subsonically expanding and rising radio lobe bubbles
which was developed by Churazov et al. (2000, 2001a).

First of all we check the first requirement concerning
the overall energetics. The interacting radio lobe bubble
model allows us to estimate the output of the AGN in form of kinetic
energy of the relativistic plasma in the jets for typical 
time scales of several $10^7$ yr. The energy estimate relies
on a comparison of the bubble inflation and buoyant rise time
as described in Churazov et al. (2000) and requires as observational
input parameters the bubble size, $r_B$, the ambient pressure,
$P_{th}$, and the Keplerain velocity at the bubble radius in the cluster,
$v_K$. Then an order of magnitude estimate of the 
energy output of the AGN can be obtained by the formula

   \begin{table}
      \caption{Estimated energy output from the central AGN in M87,
NGC1275 and the central galaxy 
of the Hydra A cluster. The input parameters are the bubble radius, $r_B$,
the ambient pressure, $P_{th}$, and the Keplerian velocity at
the bubble location, $v_K$.}
         \label{Tempx}
      \[
         \begin{array}{lllll}
            \hline
            \noalign{\smallskip}
 {\rm system}& r_B & P_{th}  & v_K & L_{kin} \\
            \noalign{\smallskip}
            \hline
            \noalign{\smallskip}
{\rm M87} & 8 {\rm kpc} & 10^{-10} {\rm erg~ cm}^{-3} & 460~ {\rm km~ s}^{-1} 
& 1.2 10^{44} {\rm erg~ s}^{-1} \\
{\rm NGC1275} & 15 {\rm kpc} & 2 10^{-10} {\rm erg~ cm}^{-3} 
& 600~ {\rm km~ s}^{-1} & 1~ 10^{45} {\rm erg~ s}^{-1} \\
{\rm Hydra~ A} & 15 {\rm kpc} & 2.8  10^{-10} {\rm erg~ cm}^{-3} 
& 550~ {\rm km~ s}^{-1} & 2~ 10^{45} {\rm erg~ s}^{-1} \\
 
            \noalign{\smallskip}
            \hline
         \end{array}
      \]
\label{tab1}
   \end{table}

\begin{equation}
L_{kin} = 10^{45} \left({r_B \over 13~ {\rm\small kpc} }\right)~   
\left({P_{th} \over 2\cdot 10^{10} {\rm\small erg~ cm}^{-3} }\right)~
\end{equation}

\begin{equation} 
\times \left({v_K \over 700~ {\rm\small km~ s}^{-1} }\right)^{-2} {\rm erg~ s}^{-1}
\end{equation}

Table 1 shows the input parameters and the resulting output energies
for the case of M87-Virgo, Perseus, and Hydra A. The results for 
Perseus can also be found in Churazov et al. (2000). For M87 we have
used the parameters for the inner bubble taken from the X-ray
deficit observed in the jet region of M87 as shown by B\"ohringer
et al. (1995).  Similar estimates were found
by Churazov et al. (2001a) from a comparison of the appearance
of the rising bubbles with the hydrodynamic simulations. 
The input parameters for
Hydra A are taken from McNamara et al. (2000). In these estimates
we assume that the bubbles are on the verge to rise, thus the observed
values are strictly to be taken as lower limits. Since the radial
expansion velocity of the bubbles decreases rapidly with time
we will most probably observe the bubbles near to their largest radius
and thus the true values are most probably close to these lower limits. 

These values
for the energy input have to be compared with the energy loss in the 
cooling flow, which is of the order of $10^{43}$ erg s$^{-1}$
for M87 and about  $10^{44}$ erg s$^{-1}$ for the Perseus cooling flow.
Thus in these cases the first requirement is met, the energy input
is larger than the radiation losses in the cooling flow for Perseus
and the M87 halo for at least about the last $10^8$ yr. 

We have, however, evidence that this energy input continued 
for a longer time with evidence given by the outer radio halo
around M87 with an outer radius of 35 - 40 kpc (e.g. Kassim et al.
1993, Rottmann et al. 1996). Owen et al. (2000) give a 
detailed physical account of the halo and model the energy input 
into it. They estimate the total current energy content in the halo
in form of relativistic plasma to $3 \cdot 10^{59}$ erg and
the power input for a lifetime of about $10^8$ years, which is also
close to the lifetime of the synchrotron emitting electrons,
to the order of $10^{44}$ erg s$^{-1}$. The power currently 
supplied by the inner jets is also of the order of 
$10^{44}$ erg s$^{-1}$ according to the model by Owen et al. (2000).
Thus, even though the sharp boundary of several features at different
scales within the radio halo suggest that the energy supply was 
not exactly steady and one may be able to distinguish several 
episodes of higher activity, the observations also imply, that
there was a typical mean energy input from the AGN in form
of relativistic plasma of about $10^{44}$ erg s$^{-1}$ for at least
about  $10^8$ years.

The very characteristic sharp outer boundary of the 
outer radio halo of M87,
noted by Owen et al. (2000), which is observed similarly at three
different radio frequencies, 74 and 330 MHz (Kassim et al. 1993)
and 10.6 MHz (Rottmann et al. 1996) has some further very important
implications. Owen et al. (2000) already noted that this excludes
the feeding of the halo by single-particle diffusion which would 
not lead to a sharp edge and not with the same location at all 
frequencies. Similarly they argue that this speaks against an outflow,
since this would lead to a more rapid radial dimming of the halo
than observed. In a similar way we can argue, that it is difficult
to produce such a sharp outer boundary (which looks even a bit 
edge-brightened) in a cooling flow by 
advection and compression of an ambient magnetic field associated
with relativistic particles (as argued by Soker \& Sarazin 1990,
see also Fabian 1994). As a result of a cooling flow advection
and compression we would expect a halo that gradually increases
in brightness inwards, contrary to what is observed.

The best explanation of the radio halo morphology is provided
by the rising radio bubble picture of Churazov et al. (2000, 2001a)
in which the halo is filled with buoyantly uplifted bubbles which
mix with ambient gas until they are loaded enough to find a 
hydrostatic equilibrium position in the M87 halo. The typical
hights that can be reached as calculated and simulated by 
Churazov et al. are in very plausible agreement with the observed size
of the outer radio halo. (This model depends on the assumptions 
made for the mixing of the thermal and relativistic plasma -
if contrary to the assumptions the mixing is not tight, the
relativistic bubbles could subsequently leave the cooling flow
region.) 

In summary, we find a radio structure providing evidence for a 
power input from the central AGN into the halo region 
of the order of about ten times
the radiative energy loss rate over at least about $10^8$ years
(for this representative example of M87).
The energy input could therefore balance the heating for at least
about $10^9$ years if all the input energy is used to heat the 
cooling flow region. The observation of active AGN in the centers of cooling 
flows is a very common phenomenon. Burns (1990) and Ball et al. (1993)
find in a systematic VLA study of the radio properties of cD galaxies
in cluster centers, that 71\% of the cooling flow clusters have 
radio loud cDs compared to 23\% of the non-cooling flow cluster cDs.
Therefore we can safely assume that the current
episode of activity was not the only one in the life of M87 and its
cooling flow. 

Since similar central radio halos have so far been found at least 
in a number of other nearby cooling flow clusters, e.g. Perseus
(Pedlar et al. 1990, Burns et al. 1992), A2199 (Owen \& Eilek 1998),
A133, A2052, A2626 (Rizza et al. 2000, Zhao et al. 1993), 
the scenario described above
for M87 should apply similarly to these cooling flow clusters. 

Having demonstrated that the first requirement seems to be fulfilled,
we can explore how a self-regulation mechanism might work, that adjusts
the energy supply from the AGN to the cooling losses. This mechanism
should most probably be searched for in a feeding mechanism of the
AGN by the cooling flow gas, as generally suggested in most models 
that featured AGN heating. The most simple physical situation 
would be given if simple Bondi
type of accretion from the inner cooling core region would roughly
provide the order of magnitude of power output that is observed
and required.
Using the classical formula for spherical accretion from a hot gas 
by Bondi (1952) we can obtain a very rough estimate for this number.
For the proton density in the environment of the M87 nucleus we find
an average value of about 0.1 cm$^{-3}$ over a region of about
15 arcsec radius and a temperature of about $10^7$ K from the
{\sl XMM-Newton} observations (Matsushita et al. 2001). Thus
for  a black hole mass of $3\cdot 10^9$ M$_{\odot}$ (e.g. Ford et al.
1994, Macchetto et al. 1997) we find a 
mass accretion rate of about 0.01
M$_{\odot}$ yr$^{-1}$ and an energy output of about $7 \cdot 10^{43}$
erg s$^{-1}$, where we have assumed the canonical value for the ratio
of the rest mass accretion rate to the energy output, that is the 
energy conversion efficiency. For the accretion radius,
which we approximately take as the radius where the Keplerian velocity
in the potential of the black hole is equal to the sound speed, 
we find approximately a value of 50 pc ($\sim 0.6$ arcsec). 
That we get with this very crude estimate so
close to the required energy output is extremely encouraging.
We should note
further, that the required accretion rate with an energy output of 
$10^{44}$ erg s$^{-1}$ is more than a factor of 1000 below the Eddington
accretion rate and thus no reduction effects of the spherical accretion
rate by radiation pressure have to be expected. We should further note
that in this model small changes in the temperature and density structure 
in the inner cooling core region will directly have an effect on the accretion
rate. Therefore we have all the best prospects for building a
successful self-regulated AGN-feeding and cooling flow-heating model.
This will be discussed further in the paper by Churazov et al. (2001b).

We have listed as a third requirement for the heating model the fact
that the heating must be global to the whole cooling flow region.
If the heating would be acting only on certain regions of the 
plasma, they would be heated and get buoyant while the rest of the plasma
would still feature a cooling flow. In the most extreme case this would
lead to classical convection, which is excluded by the observations of 
the observed entropy structure.      

To explore the heating process further we shall consider how the
energy is transferred from the relativistic plasma in the bubbles
to the thermal intracluster medium in the frame of the 
rising bubble model of Churazov et al. (2000, 2001a). In fact 
most of the primary energy input from the jets is transferred 
to the thermal plasma in form of
adiabatic work spent on the surrounding medium. 25\% of the 
energy is lost directly from the relativistic plasma
during the inflation of the bubbles.
Further about 45\% are transferred to the ambient medium
during the adiabatic expansion of the
bubbles if they rise from about 1 kpc to 40 kpc featuring a
pressure drop of about a factor of 10. 
Thus more than half of the energy is transferred in a global way
by $P dV$-work done on the ambient medium. 
Since the bubbles are expanding
subsonically (at least on average) this energy will be converted
into sound waves, gravity waves, and in the very plausible case
of an unsteady expansion of the bubbles into low amplitude shock
waves. How much of this energy can be deposited in the cooling flow
region and how the energy is spread in detail over the region 
remains to be explored in detail. Two points are nevertheless
noteworthy. The energy is not deposited directly in the
boundary of the bubbles, as it would be expected for supersonic
expansion. This is consistent with the observations where the
gas directly bounding the bubbles seems colder than the 
general medium in the cooling cores (Schmidt \& Fabian 2001).
Part of this energy should be dissipated in the cooling flow region,
in particular if low amplitude shock waves are involved, and we
may not need a 100\% efficiency of the heating to reduce the cooling flow
mass deposition, depending on the duty cycle of the on and off state
of the central AGN. 

Further energy input into the cooling flow region is gained from
the uplift of gas dragged along with the rising relativistic plasma
bubbles. During this interaction of the relativistic and thermal
plasma the thermal gas will also be heated by dissipation
of turbulence and reconnection
of the magnetic fields in the Kelvin-Helmholtz instability 
interaction regions.
Since in the model by Churazov et al. these rising bubbles of 
relativistic plasma and dragged ambient gas will finally reach
a hydrostatic equilibrium position where they will expand laterally,
this spreading will provide an effective energy distribution in 
the azimuthal direction helping to distribute the energy originated
from the AGN globally in the cooling flow region. 
As noted above, the working of this model depends also on the way
the thermal and relativistic plasma can mix. Again we refer
to the more detailed modeling that will be presented in the
following paper, but argue here that the prospects for fulfilling
the third condition looks good under the circumstances described here. 

The last two requirements noted, the consistency with the observed
entropy structure and the metallicity increase towards the center, will
not be met for example by a conventional heating model in which the
thermal intracluster medium is mostly heated at the center and where
a convectively unstable entropy profile that decreases with radius
is driving the convective energy transport. The rising radio bubble 
model of Churazov et al. (2000, 2001a) is clearly different from 
this situation, since here a kind of convective transport is mediated
by the relativistic plasma bubbles which uplift thermal gas without
effective heating, at least in the central region. Support for this
presumption in the case of M87 comes from the observation, 
that the gas within the 
upstreaming relativistic plasma is colder than the ambient medium
(B\"ohringer et al. 1995, Belsole et al. 2001). This is most probably
explained by the fact that the uplifted gas expands adiabatically and
cools. This is the so far best explanation at hand for the observation
of the cooler thermal gas within the radio lobes. The adiabatic cooling
effect has also been clearly found in the simulations of the rising 
bubble model by Churazov et al. (2001a). 

Concerning the metal distribution, some simple estimates show that
the abundance of the heavy elements in the central region (the inner
about 10 kpc) can be replenished by the effect of supernovae type Ia
and by stellar winds within about 2 Gyrs 
(see also Matsushita et al. 2001). Even the case of a cooling flow   
with slow enough inflow velocities is not strongly violating the 
time scale requirements for this to happen. Therefore slow
convection flows driven by the radio bubbles will most probably be 
also consistent with the abundance distribution profiles.
The available X-ray observations provide more details for the guidance
of further modeling in this direction that we plan to use for
future work.

\section{Magnetic Fields}

Large Faraday rotation measures have been found in cooling flow clusters
with values of a few 100 up to 8000 rad m$^{-2}$ (Dreher et al. 1987,
Owen et al. 1990, Burns et al. 1992, Taylor \& Perley 1993, Ge \& Owen 1993,
Taylor, Barton \& Ge 1994, Taylor et al. 2001) indicating that
strong magnetic fields exist in these cooling core regions of 
clusters. Previously these magnetic fields were attributed to the
advection and compression of magnetic flux and associated
relativistic particles in the cooling flows 
predicted to reach field strengths up to the equipartition value
(e.g. Soker \& Sarazin 1990). These signatures were also taken as
a reconfirmation of the cooling flow picture that makes the 
amplification of the magnetic fields possible (Fabian 1994).

Taking again the case of M87 as an example, the sharp outer boundary
of the outer radio lobes is not much in favour of such a model, as 
discussed above. The strong magnetic fields could alternatively
result from plasma accumulation from the rising radio lobe
bubbles. When the radio synchrotron emitting electrons have long lost
their energy the associated magnetic field will still persist. 
To observe in this case strong Faraday rotation measures poses another
constraint on the mixing model of the rising radio lobe bubbles.
The magnetic field has to mix with the thermal plasma, since
the observed rotation measure depends on the product of the magnetic 
field strength and the electron density. If such a mixing is 
possible is unclear at this stage and it might be difficult to predict
theoretically. (One could ask in a similar way if the observed about 10\%
influx of the solar wind magnetic field into the magnetically 
separated magnetosphere would have been predicted on purely 
theoretical grounds were there no observational indication that this has
to happen). Again we have to hope for observational signatures 
that could possibly guide our further modeling.

Nevertheless it is interesting to speculate somewhat further. Since 
the sharp boundary of the larger M87 radio halo is so intriguing,
and since we observe these strikingly sharp features of so-called
cold fronts in merging clusters (Markevitch et al. 2000, 
Vikhlinin et al. 2000, Mazotta et al. 2001) 
we may wonder, if the separation of the
material inside the cold front from the ambient medium is 
produced by the magnetical structure of a fossil radio halo
produced by the convective radio bubble model of the AGN heating
of a cooling core region. This suggestion would naturally 
provide a sharp boundary for the cold front structure.
Support for this suggestions would be provided if a jump in the
magnetic field strength could be detected across the cold front.

\section{Conclusions}
 
Several observational constraints have let us to the conclusion that
the mass deposition rates in galaxy cluster cooling cores are not 
as high as previously predicted. The new X-ray spectroscopic observations
with a lack of spectral signatures for the coolest gas phases expected
for cooling flows, the morphology of the X-ray surface brightness distribution
in the cooling flows, and the lower mass deposition rates indicated at
other wavelength bands than X-rays are more consistent with mass 
deposition rates reduced by one or two orders of magnitude below the previously
derived values. This can, however, only be achieved if the gas in the cooling
flow region is heated. While the cooling flow models, governed essentially
by the energy equation without a heating term, were in fact the most 
simple possible description, it seems now that we have to add a step
of complication and introduce an additional heating process. The most
promising heating model is most probably a self-regulated heating
model powered by the large energy output of the central AGN in
most cooling flows.

Most of the guidance and the support of the heating model proposed here
(based on concepts developed in Churazov et al. 2000, 2001a) 
is taken from the detailed 
observations of a cooling core region in the halo of M87 and 
to a smaller part from the observations in the Perseus cluster. 
These observations show that the central AGN produces sufficient heat
for the energy balance of the cooling flow, that the most fundamental
and classical accretion process originally proposed by Bondi (1952)
provides an elegant way of devising a self-regulated model of 
AGN heating of the cooling flow, and that most of the further 
requirements that have to be met by a heating model to be consistent
with the observations can most probably be fulfilled.

Since these ideas are mostly developed to match the conditions in
M87, it is important to extent such detailed studies to most other
nearby cooling flow clusters. Not in all cases the same model
necessarily works. In some clusters more violent, supersonically
expanding radio lobes may have to be taken into account (e.g. for Cyg A).
The study of a sample of cooling flow clusters could also provide
important insight into the duty cycle of the AGN activity and
the feedback mechanism.

In this new perspective the cooling cores of galaxy clusters become the
sites where most of the energy output of the central cluster AGN is
finally dissipated. Strong cooling flows should therefore be the
locations of AGN with the largest mass accretion rates. While in
the case of M87 with a possible current mass accretion rate of about
0.01 M$_{\odot}$ y$^{-1}$  the mass addition to the black hole
(with an estimated mass of about $3 \cdot 10^9$  M$_{\odot}$) 
is a smaller fraction of the total mass, the mass build-up may
become very important for the formation of massive black holes in
the most massive cooling flows, where mass accretion rates above 
0.1 M$_{\odot}$ y$^{-1}$ become important over cosmological times.

\begin{acknowledgements}                                                        
We like to thank Rashid Sunyaev, Andrew Fabian, and
Peter Schuecker for valuable comments.
The paper is based in some parts on observations obtained with XMM-Newton, 
an ESA science mission with instruments and contributions directly funded by
ESA Member States and the USA (NASA). The XMM-Newton project is supported
by the Bundesministerium f\"ur Bildung und Forschung, Deutsches Zentrum 
f\"ur Luft und Raumfahrt (BMBF/DLR), the Max-Planck Society and the
Haidenhain-Stiftung. 
\end{acknowledgements}

\appendix
\section{}

In this appendix we derive the formula to calculate the apparent
metallicity for a mixed phase plasma as used in section 2.2.
In the following calculations we argue in terms of the total metal
abundance as a unique value, but the derivation can directly 
be generalized to any metal element abundance distribution as long
as different metal element abundances are kept fixed for both phases
for the whole temperature range considered. 
Before we can start the calculations we need a reference for the
fraction of power radiated by metal lines (and the continuum 
contribution from metals through e.g. recombination) compared the fraction
of the continuum radiated by a plasma devoid of metals. We calculate
this reference function for the case of a plasma with solar abundances
as a function of temperature (using XSPEC) to define the function
${\cal F}_1$, which gives the metal line fraction to the total radiation
for solar metallicity plasma. This function ${\cal F}_1(T)$ is shown 
in Fig.~\ref{fig4}. The function can be transformed to calculate the
metal line fraction for any other metallicity value, e.g. the 
metallicity of the enriched phase of the example in section 2.2. by
the formula

\begin{equation}
{\cal F}_2 = {z_2 {\cal F}_1 \over z_2 {\cal F}_1~ + ~(1 -{\cal F}_1)} .
\end{equation}

The apparent metallicity, $z_{obs}$, is then equal to the ratio of
the observed radiation power in metal lines compared to the metal devoid
continuum in relation to the same parameters for the solar metallicity
plasma

\begin{equation}
z_{obs} = {P_{lines} \over P_{cont.}} \times { 1- {\cal F}_1 \over
{\cal F}_1} \\
   = {f~ {\cal F}_2 \over f~ (1-{\cal F}_2)~ + ~(1-f)}  { 1- {\cal F}_1 \over
{\cal F}_1} ,
\end{equation}

where $f$ is the mass fraction of the enriched plasma phase. With 
some algebra  (using the fact that $z_2 = z f^{-1}$) 
this can be transformed into 

\begin{equation}
z_{obs} = {z~(1- {\cal F}_1) \over 1~ +~ {\cal F}_1~(zf^{-1} - 1 - z)} .
\end{equation}

This function is shown in Fig.~\ref{fig5}. The dramatic decrease
of the apparent metallicity is not only caused by the fact that
the metallicity of phase 2 goes into saturation but also by the fact
the phase 1 is forced to radiate purely by continuum emission.

\end{document}